# AI-Enabled System for Efficient and Effective Cyber Incident Detection and Response in Cloud Environments

Mohammed A. M. Farzaan, Mohamed Chahine Ghanem *, Ayman El-Hajjar, Deepthi N. Ratnayake

*Abstract*—The escalating sophistication and volume of cyber threats in cloud environments necessitate a paradigm shift in strategies. Recognising the need for an automated and precise response to cyber threats, this research explores the application of Artificial Intelligence (AI) and Machine Learning (ML) and proposes an AI-powered cyber incident response system for cloud environments. The proposed system, tested in three major datasets, namely NSL-KDD, UNSW-NB15, CIC-IDS-2017, encompassing Network Traffic Classification, Web Intrusion Detection, and post-incident Malware Analysis (built as a Flask application), achieves seamless integration across platforms like Google Cloud and Microsoft Azure. The Random Forest model demonstrated accuracies of 90%, 75%, and 99% on the respective datasets for network traffic classification, and 96% for the malware analysis application, while the neural network model for the malware analysis system achieved 99% accuracy. Our research highlights the strengths of AI-powered cyber security, with Random Forest excelling at classifying cyber threats, offering an efficient and robust solution. Deep learning models significantly improve accuracy, and their resource demands can be managed using cloud-based Graphics Processing Units (GPUs) and Tensor Processing Units (TPUs). Cloud environments themselves provide a ideal platform for hosting these AI/ML systems, while container technology ensures both efficiency and scalability. These findings demonstrate the contribution of the AI-led system in guaranteeing a robust and scalable cyber incident response solution in the cloud.

## I. INTRODUCTION

In recent years, the proliferation of cyber attacks targeting organisations across various industries has reiterated the critical need for robust incident response capabilities. According to the UK government's Cybersecurity Breaches Survey in 2024 [1], the 2024 Report on the Cybersecurity Posture of the United States by the Office of the National Cyber Director, ENISA's Threat Landscape (ETL) Report 2024 [44], and IBM's Cost of a Data Breach Report [2], a significant percentage of businesses and charities have experienced breaches or attacks, with alarmingly low adoption rates of formal Incident Response (IR) capabilities [3]. Consequently, there is a pressing demand for organisations to invest in Incident Response capabilities to safeguard against data breaches and cyber threats [4]. Organisations with well-tested Incident Response (IR) capabilities and high levels of AI and ML integration for threat detection and response demonstrated substantially lower data breach costs, as highlighted by IBM's Cost of Data Breach 2024 report [2]. This demonstrates how essential it is for organisations to embrace AI and ML technologies to bolster their cybersecurity posture [36] .

This study investigates how AI contributes to cybersecurity and explores the potential of applying it in cloud environments to address associated challenges. It proposes a novel system leveraging AI and ML techniques to enhance cybersecurity within cloud environments. The proposed system includes three main components; a network traffic classifier, a web intrusion detection system (WIDS) and a malware analysis system.

The network classifier utilises real-time traffic capture to analyse ongoing network activity for anomalies potentially indicative of malicious behaviour. The NSL-KDD dataset [10],UNSW-NB15 [12], and CIC-IDS-2017 [11], which are widely recognised benchmarks for network traffic analysis, collectively serve as the foundational datasets for training and evaluating our classifier [33]. By effectively classifying incoming traffic in real-time based on this rich feature set, the classifier can significantly enhance network security by enabling prompt identification and mitigation of potential cyber-attacks [13].

The Web Intrusion Detection System (WIDS) focuses on detecting suspicious behaviour in web traffic to prevent unauthorised access [4]. It achieves this by extracting informative features from standard HTTP server logs. The key innovation of this design lies in its real-time deployment and distributed data collection using lightweight agents on web servers. This ensures efficient log collection and minimises the impact on individual servers [24]. Anomaly detection employs the Isolation Forest algorithm, which is effective in high-dimensional datasets commonly encountered in security applications. To reduce false positives, the application triggers alerts only when the number of detected anomalies exceeds a predefined threshold, which can be optimised by the network administrator after assessing and factorising the associated

---

* Mohammed Chahine Ghanem the corresponding author.

Mr M.A.M. Farzaan is with PRIAM CYBER AI Limited, London, UK. e-mail: ashfaaq@priam.ai

Dr M.C. Ghanem is with the Department of Computer Science, University of Liverpool, Liverpool, UK. and Cyber Security Research Centre. London Metropolitan University email: mohamed.chahine.ghanem@liverpool.ac.uk

Dr A. El-Hajjar is with Cyber Security Research Group, University of Westminster, London, UK. e-mail: a.elhajjar@my.westminster.ac.uk

Dr D. N. Ratnayake is with Cybersecurity and Computing Systems Research Group. University of Hertfordshire. Hatfield, UK. e-mail: d.ratnayake@herts.ac.uk

risks, based on the assumption that real-world attacks often involve rapid bursts of activity [6].

The Malware Analysis system streamlines the process of analysing suspicious files to determine if they are malicious [23]. It achieves this by first extracting key features such as string, Import Address Table (IAT), and callback servers from training binaries and then using them to train a model. The model adopts a combined model architecture to mitigate false positives. It uses the Random Forest model as the primary model, complemented by a secondary model, the Keras TensorFlow model. Both models were trained on a comprehensive dataset obtained from VirusTotal.com. The system follows a logical flow where uploaded files undergo initial processing and classification. If the initial model predicts a high likelihood of malicious content, the file is classified as "Malicious", and a detailed analysis report is generated. A secondary deep-learning model is invoked for precise prediction for files with an uncertain classification.

This paper represents a concerted effort to explore the practical application of AI techniques in the domain of digital forensics, with a specific focus on developing an AI-enabled Cyber Incident Investigations Framework tailored for deployment in cloud environments. By leveraging the capabilities of AI and ML, this research seeks to enhance the efficacy and efficiency of digital forensics processes, thereby enabling organisations to better detect, analyse, and mitigate cyber threats in cloud infrastructures. Through comprehensive investigations, this work delves into three distinct AI and ML applications of digital forensics: Network Traffic Classification, Web Intrusion Detection, and Malware Analysis Systems. These applications are meticulously integrated within leading cloud platforms such as Google Cloud and Microsoft Azure to facilitate forensic operations effectively.

The findings derived from this research shed light on several critical aspects of AI-driven digital forensics. Firstly, the suitability of Random Forest algorithm emerges prominently for classification tasks, demonstrating robust performance in distinguishing between various network behaviours and identifying potential threats. Furthermore, integrating deep learning models unveils new horizons in Malware Analysis, underscoring the potential for enhanced accuracy and efficacy in digital forensics tasks. Moreover, this research underscores the effectiveness and scalability of cloud environments as hosting platforms for AI and ML systems. By harnessing cloud infrastructures' computational power and flexibility, organisations can significantly enhance their digital forensics capabilities, thereby overcoming the constraints of traditional on-premises solutions.

Additionally, the exploration of container technology underscores its pivotal role in facilitating the deployment and scalability of AI and ML-driven digital forensics systems within cloud environments. The agility and resource efficiency offered by containerisation presents compelling advantages for organisations seeking to streamline their forensic operations and adapt dynamically to evolving cyber threats. In conclusion, this research presents a novel and pragmatic approach to combating cybercrime in cloud environments, leveraging the synergistic potential of Artificial Intelligence and cloud resources. By bridging the gap between cutting-edge AI technologies and the demand for digital forensics, the proposed AI-led Digital forensics and incident response (DFIR) system represents a crucial step towards fortifying organizational resilience against cyber threats in the digital age.

## II. RESEARCH QUESTIONS AND CONTRIBUTION

This research represents an important contribution to the cybersecurity domain, as well as a complete proposal of an AI-enabled cyber incident response system tailored for cloud environments. Unlike traditional systems, the proposed system capitalises on the strengths of ML models such as Random Forest and Deep Learning to enhance detection accuracy and efficiency during incident response within the cloud infrastructures; it has provided real-time analysis and classification of cyber attacks with promising results, which vary from Network Traffic Classification. Furthermore, the innovative use of container technology, which assures scalability and operational efficiency, allows handling current and further growth in the rapidly emerging sophistication level of these cyber threats in a digital world. Therefore, this contribution can underpin AI's potential to revolutionise cybersecurity and establish a robust, scalable framework to ensure its adoption across different cloud platforms—a standard-setting incident response methodology. In this research, our research questions are formulated as follows:

**RQ1** How would the integration of AI within cyber incident response systems precisely enhance detection and response capabilities against emerging cyber threats in cloud environments?

**RQ2** How would a unified AI-led system that encompasses a traffic classifier, malware analysis, and web intrusion detection enhance the effectiveness and efficiency of cyber incident investigations within cloud settings?

**RQ3** How do cloud platforms like Google Cloud and Microsoft Azure actuate scalability and versatility in the deployment of AI-led systems, and what is the contribution of a T-pot [32] in model development and an ELK Stack in log gathering and visualisation in proactive threat detection?

## III. RELATED WORK

This section synthesises research related to digital forensics and incident response systems in cloud environments and the integration of Artificial Intelligence (AI) and Machine Learning (ML) within these domains. The selected papers shed light on various methodologies, frameworks, and technologies to enhance cyber forensic capabilities and address emerging challenges in cloud computing security.

### A. Incident Response and Investigation in Cloud Environments

Several works have been done to address incident detection and response in Cloud environments; Stelly and Roussev [18] introduce SCARF, a container-based software framework designed to enable digital forensic processing at a cloud scale [15]. The paper advocates for leveraging containers as a solution to address critical issues in digital forensics, offering practical insights into the integration capabilities and performance considerations of the solution. However, the absence of experiments in cloud environments limits the assessment of SCARF's full potential. Hemdan and Manjaiah [20] presented a Cloud Forensics Investigation model Centred around digital forensics as a Service (DFaaS), emphasising the deployment of a forensics Server within cloud service providers' infrastructures. While the proposed model's performance and features look promising, its reliance on proprietary cloud environments restricts its applicability to public cloud deployments. Dykstra and Sherman [21] introduced FROST, a trusted digital forensics tool designed specifically for the OpenStack cloud computing platform. An essential and noteworthy feature of FROST is its focus on evidence integrity; FROST enables the reliable acquisition of virtual disks and API logs. However, its compatibility limited to OpenStack platforms presents challenges for investigations spanning diverse cloud infrastructures. Edington and Kishore [30] proposed a comprehensive forensics framework for cloud computing featuring a central forensics server and an external forensics monitoring plane. While the framework addresses key challenges in cloud forensics, its on-premise resource approach and lack of deployment in actual cloud environments necessitate further validation. Patrascu and Patriciu [31] suggest a secure framework focused on monitoring user activity in cloud environments, with a modular architecture tailored for KVM virtualisation technology [40]. Although the framework offers insights into securing cloud environments, its narrow focus on KVM may limit its applicability in heterogeneous cloud infrastructures.

### B. AI and ML in Digital Forensics and Incident Response

Different research and proposals exist on integrating AI and ML techniques in the DFIR process. Zewdie and Temechu [22] proposed a hybrid Machine Learning (ML) approach for anomaly detection in IoT and cloud environments. Their approach utilises Convolutional Neural Networks (CNN) and Support Vector Machines (SVM) to address security threats. The authors also acknowledged the limitations of traditional CNNs on small datasets and proposed the use of Bayesian CNNs for improved performance. They emphasise the importance of large datasets for training and suggest using datasets from sources like CIADA and Packt. Their solution involves data pre-processing, feature extraction using CNN, and classification using SVM, with the possibility of incorporating entropy-based anomaly detection for further improvement. While this research offers a promising direction with its exploration of Bayesian CNN, it would benefit from addressing how the chosen ML algorithms handle zero-day attacks and the computational demands of complex models on resource-constrained IoT devices. Furthermore, the paper lacks details on the evaluation methodology used to assess the effectiveness of the proposed solution. Irina Baptista et al. [39] proposed a new approach to malware detection using machine learning and a unique method of visualizing malware as images. While the reported accuracy for specific file types, such as PDF and DOC, appears promising, a thorough critical analysis and evaluation is still required. Firstly, the effectiveness against a wider range of malware formats beyond PDFs and DOCs is unclear. Secondly, the generalisability of the self-organizing neural network for unknown malware requires more exploration. Finally, the computational cost of image visualisation for real-time applications must be addressed. Overall, the approach holds merit, but broader testing and efficiency analysis is needed for a more comprehensive evaluation [23].

Al Balushi et al. [35] addressed the growing importance of machine learning (ML) in digital forensics, highlighting its potential to streamline investigations overwhelmed by vast amounts of digital evidence. Their paper investigated how ML techniques can automate tasks, improve accuracy, and expedite the forensic process, all while using different algorithms suitable for different forensic scenarios. However, a deeper dive into specific algorithms, implementation mechanisms and their strengths and weaknesses for different tasks would strengthen the analysis. Du et al. [34] investigated the application of artificial intelligence (AI) in digital forensics, emphasising its potential to tackle the backlog of cases caused by the ever-increasing volume of digital evidence. They explored how AI-based tools can automate evidence processing, thereby expediting the investigation process and increasing case throughput. The paper discussed challenges and future directions for AI in various digital forensic domains. However, a more detailed analysis of the specific AI techniques and their limitations in different forensic tasks could offer deeper insights.

Qadir, et.al [25] highlighted the crucial role of machine learning in addressing challenges in digital forensics, proposing applications such as link analysis and fraud detection. Despite its insightful analysis, the paper lacked empirical validation of the proposed techniques. Additionally, it overlooked potential drawbacks associated with using machine learning in this context, such as the substantial amount of training data required and the possibility of bias within the algorithms. Overall, the paper provides a springboard for exploring the potential of machine learning in digital forensics. Hilmand et al. [26] conducted a survey study on the application of ML in digital forensics, offering insights into various algorithms employed for tasks such as access controls and image distortion detection. The authors discussed various applications of ML in the field without delving into the specific strengths and weaknesses of each application. Additionally, the paper did not address the potential drawbacks of using ML, such as algorithm overheads and inherent biases.

Rughani [29] proposed a digital forensics framework that leverages artificial intelligence to enhance tool performance and minimise user interaction. However, it remains unclear how the framework would address the handling of entirely new types of cybercrime that are not included in its training data. While the suggested framework shows potential as a viable solution, it still needs an in-depth evaluation and validation of the results it claims to achieve.

Dunsin et al. [28] developed a multi-agent framework for digital investigations, showcasing reduced time for evidence file integrity checks. Despite promising results, the framework would benefit from validation in diverse cloud environments. In another study, Dunsin et al. [16] thoroughly examined AI and ML applications in digital forensics, summarising contributions, drawbacks, and impacts of existing research. The reviewed literature showcased the growing significance of AI and ML in enhancing digital forensics and incident response capabilities while highlighting the need for empirical validations and practical implementations to realise their full potential in cloud environments. Table I summarise the most relevant related works and provide a comparison with our proposed system, demonstrating its unique advantages . Unlike other related works, our system integrates a honeypot environment, enabling proactive threat detection. Additionally, the system's versatility is demonstrated through its ability to operate on multiple public cloud platforms and process diverse data sources. The incorporation of SIEM visualization provides a holistic view of security events, enhancing overall system effectiveness. By combining these innovative features, the proposed system presents a more robust and adaptable solution for addressing contemporary security challenges [17].

TABLE I: Summary of related works.

| Reference | Data Source | Technique Used | Approach |
|---|---|---|---|
| Stelly & Roussev. [19] | Experimental Data | (1) Containerisation is used to encapsulate individual executable modules (2) ExifTool and OpennSFW are used as worker modules | Propose a container-based software framework that integrates existing forensics tools into a processing pipeline as worker modules. |
| Nanda & Hansen. [27] | Cloud Resources | (1) Forensics as a Service (2) VM snapshots | Implement a Forensic as a Service (FaaS) solution, enabling digital forensics to be conducted efficiently through a cloud-based Forensic Server. |
| Dykstra & Sherman [21] | Virtual Disks, API logs | Openstack cloud | Suggest a set of three novel forensic tools designed for the OpenStack cloud platform, ensuring trustworthy acquisition of virtual disks, API logs, and guest firewall logs. |
| Philip et al. [37] | DNS logs | (1) Multi-agent system (2) Decentralised Model | Propose a multi-agent model for forensics investigation in domains where devices are often distributed across a wide area. |
| Rughani. [29] | Disk Images | Acquisition, Analysis and Presentation of data for forensics | Introduce a framework aimed at optimising speed and performance in investigating cyber crimes and minimizing user interactions. . |
| Irina Baptista et al. [39] | Malicious and Benign files. | (1) Malware detection based on binary visualization. (2) Neural Networks. | Describe a new approach to malware detection that combines machine learning with a creative method of visualizing malware as images. |
| Temechu et al. [22] | Log files from CAIDA and Packt. | Data pre-processing, feature extraction using CNNs, and classification using SVMs | Suggest a hybrid Machine Learning (ML) approach for anomaly detection in IoT and cloud environments using Convolutional Neural Networks (CNNs) and Support Vector Machines (SVMs) to address security threats. |
| **Our Work** | Network Traffic, HTTP Server logs, .exe files | (1) Real-Time feature Engineering for Classification. (2) Docker containers and Kubernetes in cloud environments (3) TensorFlow deep learning model to reduce false positives. | Propose and evaluate a system with multiple applications deployed to defend against cyber threats and respond to incidents. The system can interact with large amounts of data by scaling and predicting with higher accuracy. |

## IV. Methodology and Implementation

In this section, we present the methodology employed in this research and outline the systematic approach utilised to achieve the study's objectives. This section covers the design, development and deployment of the system in detail.

### A. System Design and Development

*1) Overall System:* Our research proposes a novel AI-powered system with a three-tier architecture for efficient cyber threat detection and investigation. This architecture leverages containerisation technology to isolate and deploy various functionalities across three distinct environments: Production, Honeypot, and DFIR, as illustrated in Figure 3. The production environment securely hosts critical infrastructure that the customer needs, ensuring production data's integrity, availability, and confidentiality. It also securely mirrors network traffic to the DFIR Environment VPC for analysis by AI models. The Honeypot environment, a core component of our system's innovation, utilises a T-Pot honeypot to strategically attract and deceive attackers [32]. This deception facilitates the collection of valuable training data for our continuously learning AI models.

The DFIR environment acts as the central hub for analysis. It houses a suite of security applications consisting of trained models that will perform predictions on new data points, including a network attack classifier that performs real-time classification on network traffic. Additionally, a Web Intrusion Detection System (WIDS) analyses web server logs collected from the production environment for anomalies. Furthermore, a storage bucket monitor in the DFIR environment leverages the Malware Analysis system (hosted on a separate subnet) to analyse suspicious files and perform static analysis. Subnet 3, considered the nerve centre, hosts the ELK stack (Elasticsearch, Logstash, and Kibana) for centralised storage and analysis of logs generated by the ML models. These logs enriched with insights from production and honeypot environments, empower analysts to identify patterns and anomalies that might indicate potential threats. The final subnet, Research and Development, acts as a bridge between the honeypot environment and the system. Labelled training data from the honeypot and the computing power provided by the cloud facilitate a continuous model training and deployment pipeline, ensuring our AI models stay up-to-date with evolving threats. This research explores different approaches and methods to address security problems by utilising AI/ML as a core defence mechanism. Our proposed system architecture, which was initially deployed on the Google Cloud platform as depicted in Figure 1, contributes to this goal by enabling efficient data flow and promoting a complete end-to-end AI system framework.

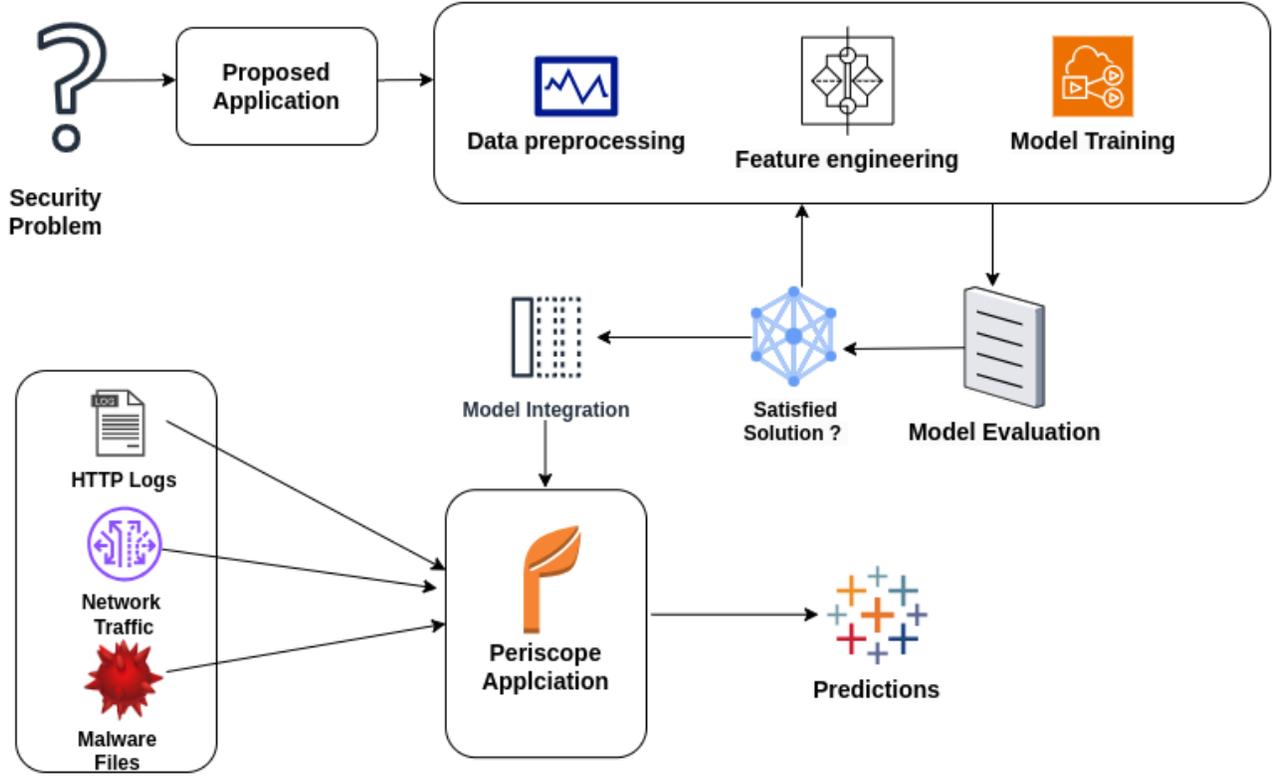

Fig. 1: Framework for Data Management and Analysis

Figure 1 depicts the proposed framework for data analysis employed in our end-to-end AI system construction, which served as the guiding structure for this research. This ten-step process, beginning with defining the business problem and culminating in the deployment of the trained model, was utilised to develop the AI system presented here. Following data selection and collection, the framework emphasised data pre-processing and feature engineering to prepare the information for model training and evaluation. An iterative loop was adopted, where model performance was assessed and potentially necessitated revisiting earlier stages in the framework for refinement. In the subsequent subsections, we also investigate various techniques for deploying AI applications within a secure and efficient architecture. Table VIII summarises the specific security problems addressed and the corresponding algorithms employed.

The data flow diagram presented in Figure 2 shows the connectivity and multiple pipelines implemented in the system. The system's design allows for centralised management through cloud infrastructure, enabling efficient monitoring and analysis. This architecture supports ongoing research and development, contributing to improved threat detection and response capabilities as shown in 3.

*2) The network traffic classifier:* Securing critical infrastructure is paramount, and network security plays a vital role in this endeavour. Attackers often exploit vulnerabilities within network systems, making network traffic analysis a crucial tool for defence. Network traffic is a rich data source containing valuable information about ongoing network activity illustrated in Table II. To effectively identify and mitigate potential threats, we propose the development of a network traffic classifier.

TABLE II: Performance Comparison of Algorithms for Traffic Classification

| Algorithm | NSL-KDD | NB15 | CIC-IDS |
|---|---|---|---|
| Random Forest | 90.92% | 75.64% | 99.82% |
| Logistic Regression | 85.75% | 63.16% | 96.2% |
| Decision Tree | 87.13% | 73.95% | 99.76% |
| KNN | 85.34% | 60.33% | 99.22% |
| Naive Bayes | 51.58% | 41.98% | 69.60% |

Algorithm 1 illustrates the functioning of Network Traffic Classification. This classifier will leverage real-time network traffic capture, allowing for online analysis of network activity.

NSL-KDD is a classic benchmark dataset with 41 features and 38 attack categories, capturing both connection-based and content-based attributes. It remains a valuable resource for detecting traditional intrusion types such as DoS, R2L, U2R, and Probe attacks. CIC-IDS 2017 [11] represents modern, real-world traffic with 80 features covering both flow-based and content-based characteristics. It includes contemporary attack types such as Brute Force, DDoS, and Infiltration, making it highly relevant for evaluating current threats. UNSW-NB15 [12] introduces 49 features that reflect a hybrid testbed environment, capturing complex attacks like Fuzzing, Backdoors, and Shellcode. This dataset allows the classifier to handle sophisticated, next-generation cyber threats with feature categories described in Table III. By training the classifier on each dataset separately, we assess its performance and adaptability across a diverse range of attack vectors and traffic environments, ensuring a comprehensive evaluation of its capabilities as detailed in Algorithm 2.

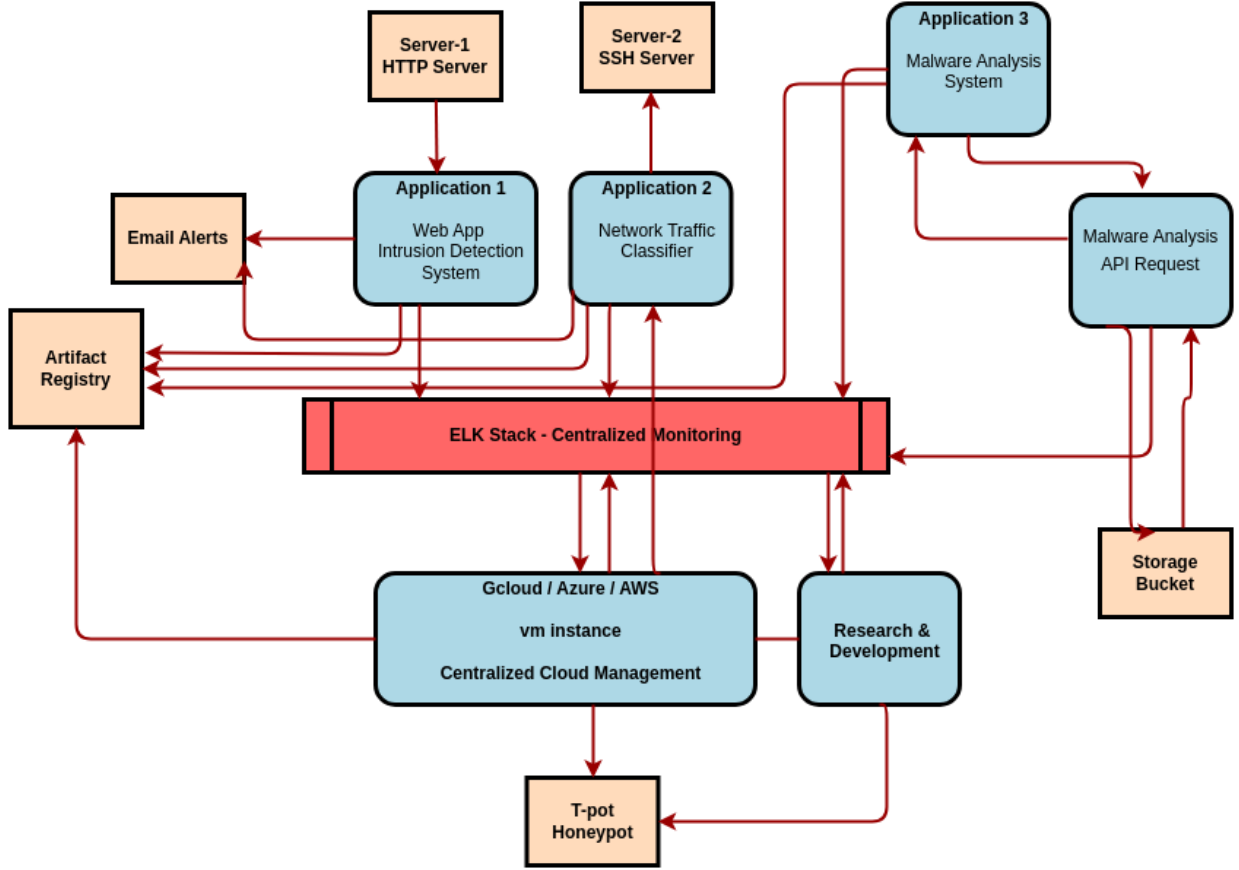

Fig. 2: Proposed System Data Flow Diagram

TABLE III: NSL-KDD Dataset Feature categories with general descriptions

| Feature category | Features | Description |
|---|---|---|
| Basic Connection | duration, protocol_type, service, flag, src_bytes, dst_bytes, land | Basic network connection attributes. |
| Host-Based | wrong_fragment, urgent, hot, num_failed_logins, logged_in, num_compromised, root_shell, su_attempted, num_root, num_file_creations, num_shells, num_access_files, num_outbound_cmds, is_host_login, is_guest_login | Characteristics related to the behavior of the host system during the connection. |
| Traffic | count, srv_count, serror_rate, srv_serror_rate, rerror_rate, srv_rerror_rate, same_srv_rate, diff_srv_rate, srv_diff_host_rate, dst_host_count, dst_host_srv_count, dst_host_same_srv_rate, dst_host_diff_srv_rate, dst_host_same_src_port_rate, dst_host_srv_diff_host_rate, dst_host_serror_rate, dst_host_srv_serror_rate, dst_host_rerror_rate, dst_host_srv_rerror_rate | Statistical features and connection rates between hosts and services. |

By effectively classifying incoming network traffic in real-time based on this rich feature set, our classifier can significantly enhance network security by enabling prompt identification and mitigation of potential cyber-attacks. Unlike existing monolithic solutions, this novel approach to network traffic classification by containerizing key components into distinct engines: the Capture Engine, Feature Engine, and Model Classifier, offers modularity and flexibility, enabling users to independently scale and deploy each component according to their specific requirements. This approach streamlines the feature extraction by automating feature engineering from PCAP files, facilitating seamless integration with diverse network environments. The network traffic classifier application encompasses a comprehensive and modularised design workflow tailored to handle the complexities of real-time network traffic analysis within cloud environments. The logical design unfolds through three interconnected components, each contributing crucial functionalities to the overall system. Firstly, the packet capture engine serves as the foundational component, leveraging the Scapy library to capture network packets continuously. Operating within its designated container, this engine listens on specified interfaces, intercepts network traffic, and stores captured packets in PCAP files for subsequent analysis. Secondly, the packet analysis module, encapsulated within another container, reads the captured PCAP files, extracts connection-based statistics, and transforms them into structured datasets

**Algorithm 1** Network Traffic Classification using Random Forest

**Require:** Training data: Network traffic dataset $D = \{(\mathbf{x}_1, y_1), (\mathbf{x}_2, y_2), ..., (\mathbf{x}_N, y_N)\}$
1: where $\mathbf{x}_i$ is a feature vector and $y_i$ is the attack type label.

**Ensure:** Traffic Classification model $M$

      **Pre-process data**
2: Read data from CSV files.
3: Drop irrelevant features (e.g., flags, protocols, services).
4: Separate features ($\mathbf{X}$) and labels ($\mathbf{y}$).
5: Split data into training and testing sets: $(X_{train}, y_{train}), (X_{test}, y_{test})$.
6: Standardize features using StandardScaler.

      **Build Random Forest model** $M$
7: Define a Random Forest classifier with a desired number of estimators (e.g., 100).
8: Set random state for reproducibility (e.g., 42).
9: Train model $M$ on $(X_{train}, y_{train})$.
10: Evaluate model $M$ on $(X_{test}, y_{test})$ using metrics (e.g., accuracy, classification report).
11: **return** Trained Traffic Classification model $M$

---

**Algorithm 2** Cloud-based Malware Classification with LSTM

**Require:** Training data: executable file $(d_1, l_1), (d_2, l_2), ..., (d_N, l_N)$
1: where $d_i$ is a file and $l_i$ is its label. Number of classes: C
2: Maximum vocabulary size: V
3: Maximum sequence length: T

**Ensure:** Trained malware classification model $M$
4: Preprocess text data:
5: Tokenize files $(d_i)$ into sequences of integers $(w_i^1, w_i^2, ..., w_i^{|d_i|})$.
6: Pad sequences to uniform length $(w_i^1, w_i^2, ..., w_i^T)$.
7: Encode labels: Convert text labels $(l_i)$ to numerical labels $(y_i)$.
8: Split data into training and testing sets: $(X_{train}, y_{train}), (X_{test}, y_{test})$.
9: Build LSTM model $M$:
10: Define a sequential model.
11: Add Embedding layer with vocabulary size V and embedding dimension E.
12: Add LSTM layer with hidden units H.
13: Add Dense layer with C output units and sigmoid activation for C classes.
14: Compile model $M$:
15: Set optimizer (e.g., Adam).
16: Set loss function (binary_crossentropy for binary classification, categorical_crossentropy for multi-class).
17: Set metrics (e.g., accuracy).
18: Train model $M$ on $(X_{train}, y_{train})$ for a desired number of epochs.
19: Evaluate model $M$ on $(X_{test}, y_{test})$ using metrics (e.g., accuracy).
20: **return** Trained malware classification model $M$

---

suitable for predictive modelling. Utilising the pre-trained machine learning model during the development phase, it predicts predefined attack labels. Finally, integrated into a separate container, the alerting mechanism monitors prediction outcomes and triggers alerts in real-time upon detecting anomalous network behaviour. These alerts serve as actionable insights for security analysts, enabling timely responses to potential threats. This containerised architecture ensures scalability, flexibility, and isolation, facilitating seamless deployment, management, and scalability of the network traffic classifier application within diverse computing environments.

*3) Web Intrusion Detection System:* This research implements a web intrusion detection system leveraging real-time anomaly detection techniques. Anomaly detection, in the context of network and host security, identifies unusual activities that may signify an attacker's presence. However, traditional anomaly detection often struggles with defining "normal" behaviour, leading to a high rate of false alarms. This research aims to address this challenge by employing AI for anomaly detection within the specific domain of web application security. The system focuses on extracting informative features from standard HTTP server logs.

Although these logs offer only a partial view of the network's overall traffic, they still contain key features that can be extracted, as illustrated in Table IV. Building a dependable and comprehensive feature set is essential for the effectiveness of the anomaly detection process. By examining these features, the system can detect anomalies that may signal various types of attacks, including those identified in the OWASP Top Ten [43].

The key novelty of this design lies in its real-time deployment and distributed data collection. Each web server can deploy a lightweight agent responsible for collecting and forwarding logs promptly to the Web Intrusion Detection System (WIDS). This agent-based approach ensures efficient log collection and minimises the impact on individual web servers. The WIDS utilises a shared volume accessible by the virtual machine (VM) and the deployed container. This shared volume facilitates efficient storage and access to the collected logs for real-time analysis.

Anomaly detection identifies unusual activity that deviates from established patterns of normal behaviour. Common approaches include:

- Statistical Metrics: Analysing deviations from statistical properties like mean, standard deviation, or frequency distribution.
- Unsupervised Machine Learning: Employing algorithms that learn patterns from unlabelled data to identify outliers.
- Goodness-of-Fit Tests: Evaluating how well a data sample fits a pre-defined statistical model.
- Density-Based Methods: Identifying anomalies as data points located in low-density regions of the feature space.

This application leverages the Isolation Forest algorithm due to its effectiveness in high-dimensional datasets often encountered in real-world security applications. Isolation Forest works by isolating potential anomalies by randomly partitioning the data. Instances that are easier to isolate are likely anomalies, while those requiring deeper partitioning are more likely to be normal data points. A critical aspect of anomaly detection is defining the threshold for flagging data points as anomalies.

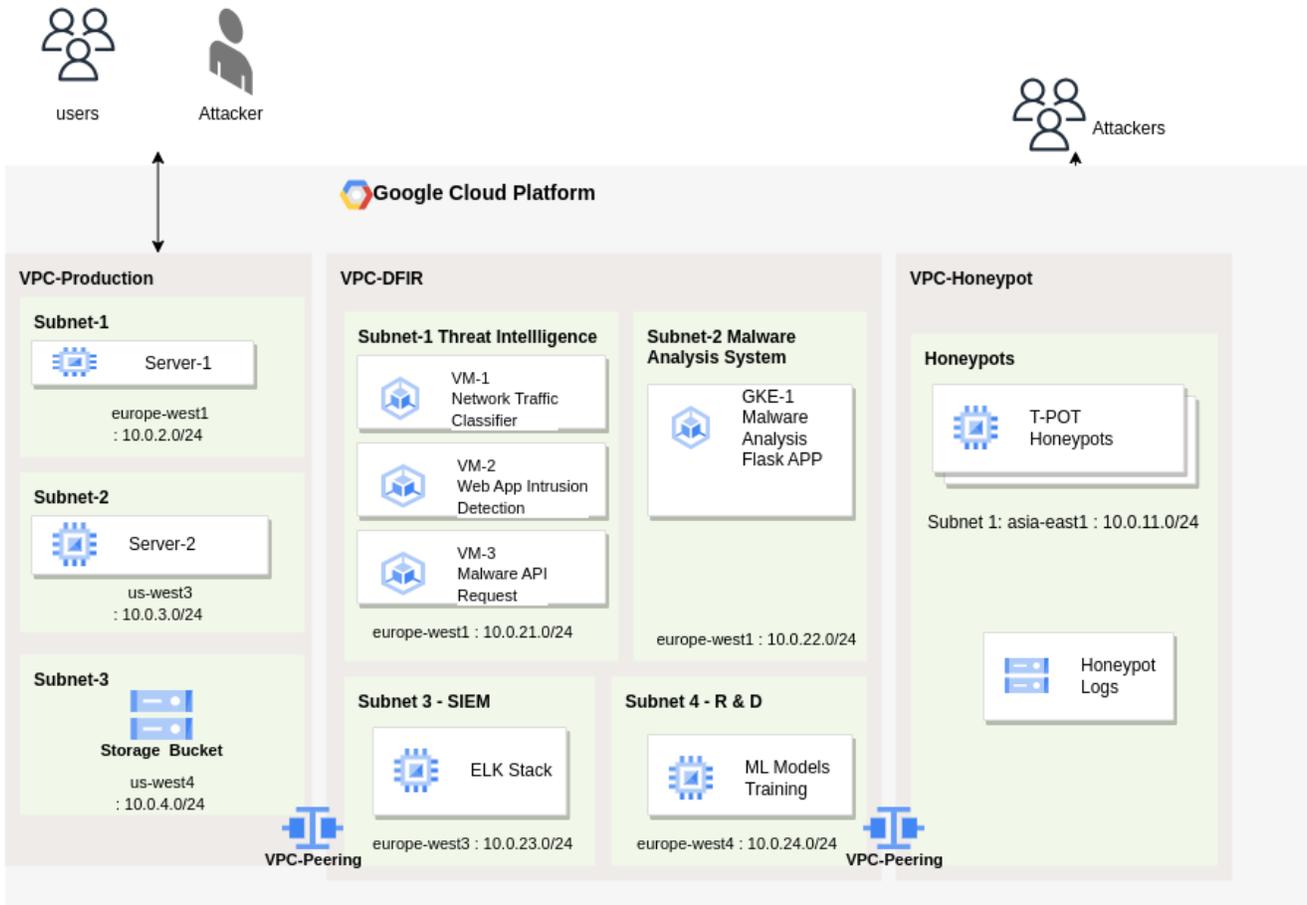

Fig. 3: System Detailed Architecture

TABLE IV: Features generated from HTTP Server logs

| Feature Name | Description |
| --- | --- |
| 1-IP-level access statistics | High frequency, periodicity, or volume by a single IP address or subnet is suspicious. |
| 2-URL string aberrations | Self-referencing paths (/./) or backreferences (/../) are frequently used in path-traversal attacks. |
| 3-Unusual referrer patterns | Page accesses with an abnormal referrer URL are often a signal of unwelcome access to an HTTP endpoint. |
| 4-Sequence of accesses to endpoints | Out-of-order access to HTTP endpoints that do not correspond to the website's logical flow |
| 5-User agent patterns | Alerts on never-before-seen user agent strings or extremely old clients which are likely spoofed. |

In this web intrusion detection system, optimizing the hyperparameters of the Isolation Forest algorithm was essential for achieving reliable anomaly detection using server log files. Initially, we set the sample size ($num\_samples$) to 1,000, which defines the number of data points used to build each isolation tree. This value was later increased to 10,000 to evaluate scalability and enhance performance, significantly improving the model's ability to detect outliers in larger datasets. The random state ($random\_state$) parameter, responsible for reproducibility, was refined through multiple tests. After experimenting with various values, it was optimised to 100, ensuring consistent model training without impacting overall performance. A novel approach was employed to determine the optimal contamination ratio by simulating specific attack scenarios (e.g., path traversal attack) and analysing the model's detection rate. This technique leverages domain knowledge about potential attack vectors to fine-tune the contamination threshold for real-world applications. We tested contamination levels of 0.001, 0.01, and 0.1. After extensive testing, a contamination ratio of 0.01 provided the best balance between anomaly detection sensitivity and minimizing false positives, making it the most effective for this system. More detailed test results are provided in the Results section. The systematic optimization of these hyperparameters led to a model that successfully balances computational efficiency and detection accuracy, tailored for real-time web intrusion detection in dynamic server environments.

Due to the inherent difficulty of acquiring perfectly representative datasets for all scenarios, continuous monitoring and adaptation are essential for robust security. This application prioritises simplicity and efficiency by utilizing a single container. The container performs the following tasks:

- Reads HTTP access logs from web servers.
- Pre-processes the log data by converting it to a pandas data frame for manipulation in Python.
- Generates features mentioned in Table IV for anomaly detection from the log data.
- Sends the extracted features to the trained Isolation Forest model for real-time anomaly detection.

To reduce false positives, the application triggers alerts only when the number of detected anomalies exceeds a predefined threshold. This approach is based on the assumption that real-world attacks often involve rapid bursts of activity, leading to a surge in detected anomalies. This design demonstrates a practical approach to deploying AI-powered anomaly detection for web intrusion detection in real time. By leveraging Isolation Forest and a domain-informed threshold determination method, the system aims to achieve efficient and accurate anomaly detection. Algorithm 3 illustrates the web intrusion detection model using the HTTP access logs.

---

**Algorithm 3** Web Server Logs Anomaly Detection using Isolation Forest

**Require:** Web server log file $L$
**Ensure:** Anomaly detection model $M$
1: **Preprocess data**:
2: Read log file $L$ into a DataFrame $D$.
3: Remove missing values from $D$.
4: **Extract features**:
5: IP-level statistics: ip_frequency, Unique_connections_count, ip_volume
6: URL aberrations: url_aberrations
7: Unusual referrer patterns: unusual_referrer
8: User-Agent analysis: user_agent_analysis (categorical)
9: Out-of-order access: out_of_order_access
10: Standardize numerical features using StandardScaler.
11: **Train Isolation Forest model**:
12: Create an IsolationForest model $M$ with:
13: Max samples: $N$ (number of samples in $D$)
14: Contamination: $c$ (estimated outlier ratio)
15: Random state: $r$ (for reproducibility)
16: Train $M$ on features $X$ in $D$.
17: **Detect anomalies**:
18: Use $M$ to predict anomaly scores $y_pred$ for new data points.
19: Mark data points with scores below a threshold as anomalies.
20: **return** Anomaly detection model $M$

---

*4) Malware Analysis System:* Malware analysis, encompassing the investigation of malicious software's functionality, purpose, origin, and potential impact, traditionally demands extensive manual effort and expertise in software internals and reverse engineering. Our research introduces a novel application streamlining this process, offering efficient and automated malware analysis capabilities.

The application development commenced with implementing the code to extract features from training binaries. Feature extraction involves identifying and collecting pertinent data from training binaries, which are then stored within a Python dictionary. Detailed information regarding the dataset utilised in the study is provided in Table V. The dataset utilised in this research, primarily sourced from VirusTotal, presents both advantages and limitations that affect its applicability in real-world scenarios. While VirusTotal offers a diverse collection of malware samples, it may be biased toward more commonly reported threats, potentially neglecting emerging or niche malware types. Additionally, the temporal relevance of the dataset is crucial, as more samples may be needed to maintain the model's effectiveness against current threats. The ratio of benign to malicious samples can impact model performance and may result in a model that over-fits malicious samples, affecting its ability to generalise well to real-world scenarios.

TABLE V: Malware Dataset

| Dataset | | Samples | |
|---|---|---|---|
| Type | Purpose | Count | Percentage |
| Benignware | Training | 694 | 70% |
| | Testing | 297 | 30% |
| Malware | Training | 299 | 70% |
| | Testing | 129 | 30% |

Regarding generalizability, the model's effectiveness may vary across different operating systems, network conditions, and user behaviours, highlighting the need for evaluation in diverse environments. Malware behaviour can vary significantly across different operating systems, application environments, and network configurations. For example, a model trained primarily on Windows malware might perform less effectively on macOS or Linux systems, where malware characteristics and user behaviour differ. To enhance practical applicability, ongoing efforts to diversify the training dataset and adapt to evolving threats are essential for maintaining robust malware detection capabilities. Incorporating a detailed analysis of the dataset, addressing its limitations, and discussing the generalizability of the model across different environments enhances the robustness of the research.

Subsequently, a model was trained using the extracted string features from the samples. In addition to string features, the malware analysis system has integrated behavioural analysis to effectively counter the stealthy tactics employed by advanced persistent threats (APTs). This includes the examination of Portable Executable (PE) header features, Import Address Table (IAT) characteristics, and the identification of callback servers. These elements play a critical role in recognising the subtle behaviours associated with sophisticated attacks. As part of future work, the malware analysis system can further enhance its detection capabilities by incorporating contextual threat intelligence about known APT groups, including their Tactics, Techniques, and Procedures (TTPs) and associated Indicators of Compromise (IoCs). This integration will enable the system to make more informed decisions based on the behavioural patterns and attributes shared by various malware samples, such as embedded IP addresses, hostnames, strings of printable characters, and graphics.

By training the detection model on malware utilised by APT

TABLE VI: Performance Comparison of Algorithms for Malware Analysis

| Algorithm | Accuracy | Precision | Recall | F1-Score |
|---|---|---|---|---|
| Random Forest | 96.71% | 94.44% | 94.44% | 94.44% |
| Support Vector Machine | 91.54% | 95.91% | 74.60% | 83.92% |
| Logistic Regression | 94.60% | 97.24% | 84.12% | 90.21% |
| Decision Tree | 88.26% | 87.25% | 70.63% | 78.07% |

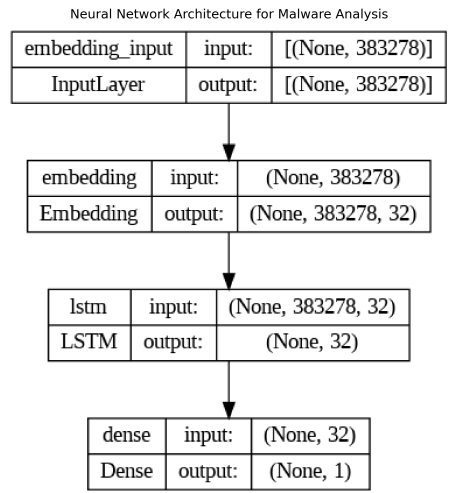

Fig. 4: Keras model parameters

groups, it can evolve beyond a binary classifier, providing a more nuanced understanding of threats. This holistic approach ensures that the malware analysis component is better equipped to address the complexities of APTs, ultimately enhancing its real-world applicability and effectiveness in defending against sophisticated cyber threats.

Various algorithms were explored during the training phase, and the most suitable one was selected based on evaluation metrics such as accuracy, precision, recall, and F1 score. To facilitate comparison, Table VI presents a summary of the performance of three different algorithms employed in the study.

The models were trained using Google Colab with T4 GPU acceleration, optimising computational efficiency. Notably, the application adopts a combined model architecture to mitigate false positives effectively. The primary model in this architecture is a random forest model complemented by a secondary Keras TensorFlow model. Both models were trained on a dataset obtained from VirusTotal.com, ensuring comprehensive coverage of malware samples. The integration of a deep learning model as the secondary component aims to reduce latency in real-time applications, enhancing the system's responsiveness as summarised in Figure 4.

- Random Forest Model: This model provides a relatively fast initial classification with good accuracy.
- Keras TensorFlow Model: This deep learning model offers additional refinement, particularly for complex malware samples.

The deep learning model employs a Long Short-Term Memory (LSTM) network, a type of Recurrent Neural Network (RNN) well-suited for analysing sequential data like text as explained in Table VII. Table VI summarises the architecture of the neural network used in the malware analysis system, detailing each layer's function, the number of neurons, and the activation functions applied. The model begins with an embedding layer that converts discrete words into dense vector representations (32 dimensions in this case). An LSTM layer with 32 hidden units then processes these sequences. Unlike the embedding layer, each hidden unit within the LSTM can be considered a "neuron" that learns and extracts features from the data as summarised in Table VII. Finally, a dense layer with a single output neuron and a sigmoid activation function performs binary classification. This single output neuron leverages the features learned by the LSTM layer to predict the maliciousness of the analysed file. The Algorithm 2 illustrates the cloud-based malware classification using the LTSM network.

The system, as illustrated in Figure 3, follows a logical flow that begins with the user uploading a file through the application's user interface. Upon submission, the uploaded file undergoes initial processing, where its features are extracted and prepared for classification. Utilising a pre-trained machine learning model, the system predicts the probability of the file being malicious. If the probability exceeds a threshold of 0.7, indicative of a high likelihood of malicious content, the system immediately classifies the file as "Malicious" and generates a detailed analysis report in PDF format. Conversely, if the probability falls between 0.5 and 0.7, the system invokes a Keras deep learning model for further binary classification. Upon classification completion, the system produces a comprehensive analysis report, facilitating informed decision-making regarding the file's security implications. For files deemed benign based on the classification results, the system provides a "Benign" classification and presents the analysis outcome to the user through a web interface. This approach combines the power of cloud computing, deep learning, and web technologies, offering a robust solution for real-time malware detection and analysis with enhanced user accessibility and efficiency.

The versatility of the application enables its utilisation in diverse scenarios. Firstly, it serves as a web service, allowing online users to upload executable files for malware analysis. The application swiftly determines the maliciousness of the files and provides results to the users. Secondly, it functions as an API, facilitating the scanning of storage buckets deployed in cloud environments. This multi-faceted approach ensures the broad applicability and practicality of the developed malware analysis solution, catering to varying user requirements and deployment environments.

### B. System Implementation

While prior research has explored the potential of AI/ML for cyber security through theoretical frameworks, there is a scarcity of studies demonstrating practical implementation. This research addresses this gap by deploying a trained AI/ML system in a real-time cybersecurity environment. This deployment attempts to bridge the theory-practice divide in this domain. The system is implemented in a public cloud environment to address the challenges in three main areas of cyber security. This real-time deployment allows for faster threat detection and response, potentially leading to improved security posture. This research deployed three AI applications in a real-world cyber security environment to bridge the gap between theoretical frameworks and practical implementation.

TABLE VII: **Neural Network Architecture**

| Layer | Function | Neurons | Activation Function |
|---|---|---|---|
| Embedding | Converts words to dense vectors | max_words (adjustable) | N/A (vector representation) |
| LSTM | Captures long-range dependencies in sequences | 32 | Sigmoid, tanh, or ReLU (default: sigmoid) |
| Dense | Performs final binary classification | 1 | Sigmoid |

- Web Intrusion Detection: A single container deployed on a virtual machine (VM) analyses web server access logs for anomalies. This simple deployment prioritises ease of maintenance for this initial application.
- Malware Analysis: A Flask application, deployed on Kubernetes Engine, classifies uploaded executable files (exe) as benign or malware. This containerised approach allows for easier scaling and updating of the model as needed.
- Network Traffic Classifier: This application, deployed using multiple Docker containers on a single VM, analyses network traffic captured as PCAP data for real-time attack detection. Containerization again facilitates model updates and simplifies deployment.

The deployment strategy leverages containerization (e.g., Docker) for several advantages:

- Scalability: Containers enable easy scaling of computational resources to meet changing demands.
- Simplified Updates: Containerised models streamline updates, reducing downtime and improving maintenance efficiency.
- Portability: A single trained model can be deployed across diverse cloud environments by pulling the container image from a central registry like Docker Hub. This simplifies multi-cloud deployments.

In the system implementation, data transmission is handled with a focus on security and efficiency. Each web server is equipped with an agent that periodically transfers HTTP server logs to its designated system container using the Secure Copy Protocol (SCP). This ensures secure, encrypted transmission of logs, protecting sensitive data during transfer. For the network traffic classifier, packet mirroring is utilised. All network traffic, including packet payloads and headers destined for the application server, is mirrored to the virtual machine (VM) running the network traffic classifier. This approach ensures comprehensive monitoring of all traffic, allowing the system to detect anomalies, security breaches, and intrusions effectively. By mirroring all traffic, the network traffic classifier can conduct thorough inspections across multiple packets, ensuring that no malicious activity is missed.

However, it is important to acknowledge the potential challenges associated with deployment. While containerization offers a robust deployment strategy, real-world implementation presents potential challenges. A critical factor is ensuring the deployed model's throughput meets the demands of real-time operation. The production system's hardware platform (CPU, GPU, memory) must provide sufficient computational resources to handle the processing demands. Careful consideration of these resource requirements is essential to avoid bottlenecks and maintain the effectiveness of the AI system in a real-time cybersecurity environment.

The deployment of the system begins with the Network Traffic Classifier, initially deployed on a single VM instance to analyse captured PCAP data in real-time. To ensure scalability and prevent potential bottlenecks as traffic volume increases, the system will implement horizontal scaling by leveraging a load balancer. This approach will distribute the network traffic across multiple VM instances, ensuring efficient and balanced processing. Vertical scaling will be employed for the Web Intrusion Detection application, with each web instance being assigned its own dedicated VM. This ensures focused resource allocation and simplified maintenance, particularly for handling the growing volume of web server access logs. Finally, the Malware Analysis application will utilise a load balancer to distribute file uploads across containers, with an additional node dedicated to API handling. Table IX presents a summary of application scalability within the cloud environment. This setup optimises scalability and performance, ensuring seamless file classification and system responsiveness even under heavy workloads illustrated in Table VIII.

Furthermore, as the demands on the AI system grow, on-premises infrastructure may struggle to keep pace. Cloud platforms offer an excellent solution for scaling computational resources to meet these growing needs. Cloud environments can also be particularly adept at deploying artificial neural networks, which often have significant computational requirements for both training and inference. cloud platforms offer additional benefits for real-world AI deployments. Cloud environments facilitate red-green deployments, allowing for seamless model updates with zero downtime. Additionally, many cloud providers offer Machine Learning specific services for training, deployment, and management of AI models. While this research utilised core cloud services for deployment demonstration, these specialised services can further streamline the process. Finally, infrastructure, such as code tools like Terraform, can automate the deployment process in cloud environments. This ensures consistent and repeatable deployments, minimising human error and configuration inconsistencies. By leveraging containerisation, cloud platforms, and infrastructure automation, this research establishes a robust and scalable framework for deploying AI systems in real-world cybersecurity applications.

12TABLE VIII: Application Scalability

| Application | Scaling Method | Load Management |
|---|---|---|
| Network Traffic Classifier | Horizontal Scaling | Load balancer distributes network traffic across multiple VMs to prevent bottlenecks. |
| Web Intrusion Detection | Vertical Scaling | Individual VMs handle anomaly detection for specific web servers, resource allocation is based on web traffic. |
| Malware Analysis System | Horizontal Scaling: | Load balancer distributes file uploads across multiple containers to ensure smooth handling of a large number of requests |

## V. Testing and Results

### A. Network Traffic Classifier

During development, various machine learning models were evaluated on the NSL-KDD, CIC-IDS 2017, and UNSW-NB15 datasets to identify the most effective approach for classifying network traffic patterns. The Random Forest algorithm emerged as the frontrunner across all three datasets, achieving accuracy rates of 90.92%, 99.82%, and 75.64%, respectively. This significantly outperformed other models, including Logistic Regression, Decision Tree, KNN, and Naive Bayes, as shown in Table II. In optimising the Random Forest classifier for network traffic classification, specific hyperparameters were carefully tuned across all three models to improve detection performance. We used Gini impurity as the criterion to measure the quality of splits, ensuring the model could effectively distinguish between classes. The minimum sample split was set to 2, allowing the trees to split nodes even with the smallest number of samples, thereby refining decision boundaries for improved accuracy. The maximum samples parameter was set to None, ensuring the models were trained on the entire training set as illustrated in Figures 5 and 6. Additionally, a random state was specified for reproducibility, ensuring consistent results across multiple runs. The NSL-KDD dataset, consisting of 125,973 training samples and 22,544 testing samples, was used with its predefined split to maintain consistency with previous studies. In the case of the CIC-IDS 2017 dataset, which contains approximately 2.8 million samples, the data was randomly divided into 80% for training (around 2.24 million samples) and 20% for testing (around 560,000 samples), ensuring a fair distribution of attack types and normal traffic across both sets. Similarly, the UNSW-NB15 dataset, with 175,341 training samples and 82,332 testing samples, was also used with its predefined split. The systematic optimisation of these hyperparameters improved the model's capacity to classify network traffic accurately, making it robust in detecting network anomalies.

Following the deployment of the Network Traffic Classifier application in cloud environments using a modular, containerized approach—featuring separate containers for packet capturing and real-time classification of incoming traffic data, pre-processing and feature engineering were performed within a single container, streamlining the workflow before classification. Attack simulations were then conducted on the application to evaluate its performance.

Table IX summarizes the results obtained from these attacks, detailing the methods used, the corresponding machine-learning models, and their performance in detecting the attacks. The results highlight the models' performance against specific attacks relevant to the datasets on which they were trained. A comprehensive evaluation was conducted to assess the system's real-world capabilities by simulating various cy-ber attacks on resources purposely made vulnerable and hosted within cloud environments. The testing was carried out using a Kali Linux environment, employing a variety of penetration testing tools to conduct attacks both remotely and within the cloud infrastructure. These attacks were executed at different stages of the cyber kill chain to evaluate the effectiveness of the trained models against specific threat vectors. Overall, the evaluation process effectively demonstrated the system's strengths and weaknesses in detecting various cyber threats, emphasising the importance of continuous refinement and training of the detection models to enhance their efficacy in real-world applications.

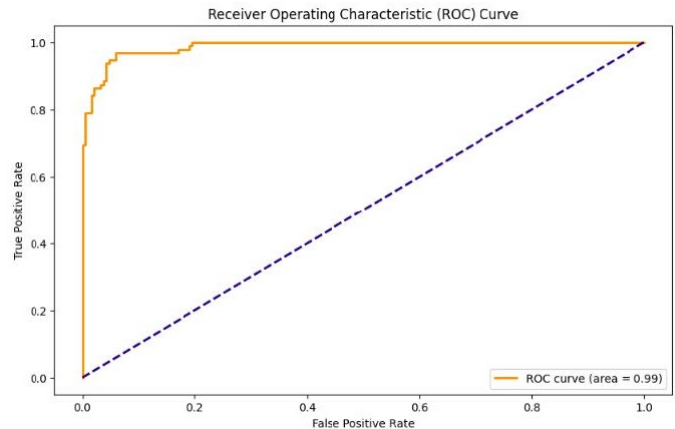

Fig. 5: Keras model ROC curve

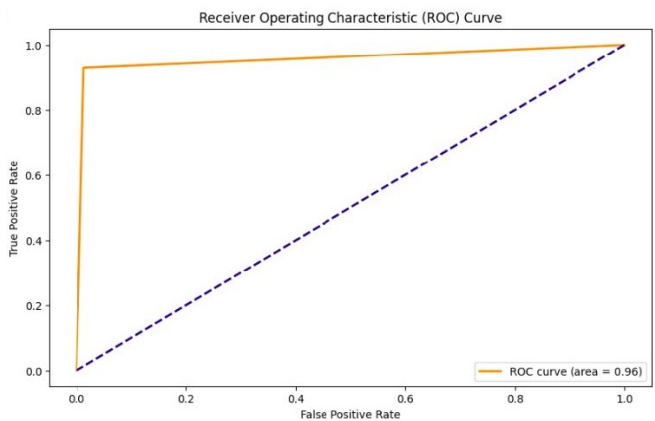

Fig. 6: Random Forest ROC Curve

### B. Web Intrusion Detection

Our research investigated the effectiveness of anomaly detection for web intrusion prevention. A key parameter, contamination, was explored to determine the model's sensitivity to anomalous traffic patterns. We evaluated different contamination levels (0.01 and 0.001) to assess the model's ability to differentiate normal website traffic from potential attacks illustrated in Figures 7 and 8.

TABLE IX: Attack Simulation Results

| Attack Type | Attack Tool/Method | Model Used | Results & Comments |
| --- | --- | --- | --- |
| Brute Force Attack | Hydra (SSH/FTP) | Random Forest - (CICIDS2017) | Low FP, High accuracy |
| Denial of Service (DoS) | Nikto (SYN Flood) | Random Forest - (NSL-KDD) | Moderate Detection, few FPs |
| Fuzzing | wfuzz (Web App Fuzzer ) | Random Forest - (UNSW-NB15) | High Detection with consistency |
| Port Scan | Nmap | Random Forest - (CICIDS2017) | Consistently detected |
| SQL Injection | sqlmap | Random Forest - (CICIDS2017) | Detected quickly, few FNs |
| Exploits | Metasploit Framework | Random Forest - (UNSW-NB15) | Lower Detection during attack |
| R2L FTP write | php web shell | Random Forest - (NSL-KDD) | Low detection, High FPs |

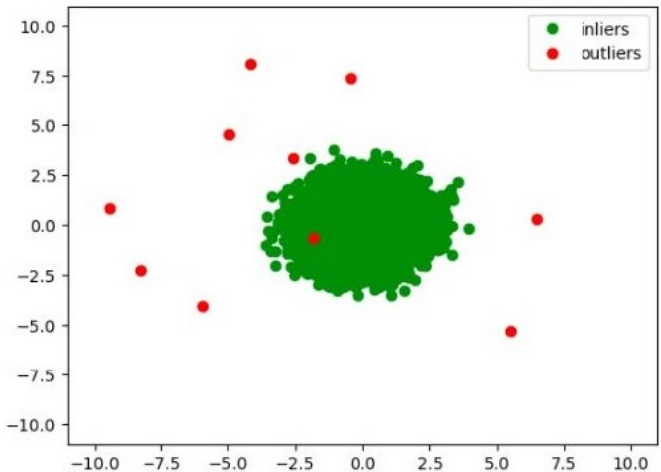

Fig. 7: Contamination 0.01

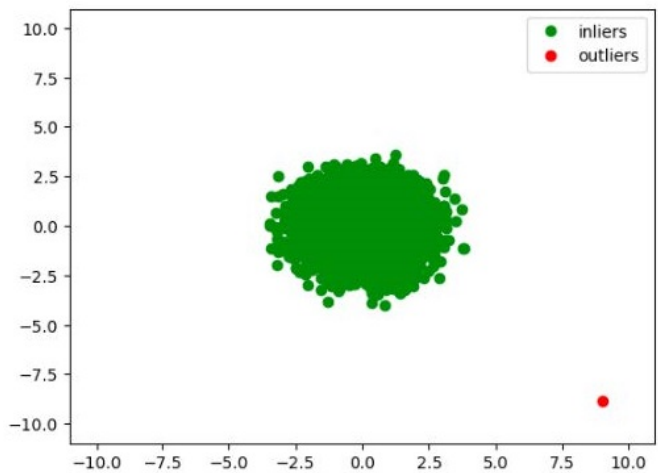

Fig. 8: Contamination 0.001

Following model training with features extracted from HTTP logs (as detailed in the methodology section), the system was integrated and deployed on the Azure cloud environment. A real-world simulation involved generating live traffic targeting a public website's IP address. The system effectively detected a simulated directory traversal attack launched using the Nikto tool on port 80. This detection triggered immediate logging and email notification, demonstrating the system's proficiency in real-time intrusion identification and response.

In addition to simulated attacks, the system underwent traffic load testing to evaluate its response under various conditions. A high volume of benign traffic was generated, during which the system maintained a high detection rate for regular activities but produced a few false positives, indicating areas where feature engineering could reduce noise in benign environments. These findings emphasise that fine-tuning the web intrusion detection model should be closely aligned with the specific application being hosted and the associated business requirements. By understanding the structure, functionality, and expected usage patterns of the application, we gain valuable insights into the types of attacks that are more likely to occur. This understanding allows us to generate tailored features that enhance the model's ability to detect these specific threats, resulting in higher accuracy and relevance in identifying malicious behaviour.

### C. Malware Analysis System

Following the performance metrics evaluation, Table V provides a summary of the algorithms tested in the malware analysis system along with their respective evaluation metrics, including accuracy, precision, recall, and F1 score. This table offers a comparative overview, highlighting the strengths and weaknesses of each algorithm in detecting malware accurately and effectively, aiding in the selection of the most suitable model for specific malware analysis needs. Additionally, Figures 5 and 6 illustrate the ROC curves for the Random Forest and Neural Network algorithms, respectively. These curves provide a visual representation of each model's true positive rate against the false positive rate, helping to assess the effectiveness of each algorithm in distinguishing between malicious and benign files. The ROC curves further validate the performance of these models, with a higher area under the curve (AUC) indicating stronger predictive capabilities.

The Random Forest algorithm exhibited an accuracy of 96%, while the Keras model achieved an even higher accuracy of 99%. Additionally, the dual model malware analysis was deployed as a Flask Application. Practical testing involved uploading an IRC Bot executable file to simulate real-world scenarios. The system accurately predicted the presence of malware and generated a detailed Static Analysis PDF report. Furthermore, the system successfully identified benign applications, such as a Minesweeper executable game file.

Further testing included automating scans on storage buckets within the cloud, enabling the system to scan and identify any malicious files stored within these environments. These scans can be configured according to administrator preferences, providing flexibility based on organizational needs. The system's API was tested by sending files to an endpoint and receiving a JSON response, detailing whether the file was classified as malicious or benign. This API integration allows the malware analysis system to be used in conjunction with other programs, enhancing interoperability and scalability across different environments.

The dual-model approach, coupled with successful practical testing, highlights the malware analysis system's capability to reliably identify and report on malicious entities while maintaining high accuracy in distinguishing benign software.

## VI. Discussion

The results obtained from the testing phase serve as crucial indicators of the overall effectiveness of the implemented system. Notably, the Random Forest algorithm used in both applications demonstrates impressive accuracy in both multi-class and binary classification tasks. Moreover, the accuracy achieved by the Keras deep learning model surpasses that of the Random Forest algorithm, reaching an impressive 99%. However, due to its high compute resource utilization during the prediction of new data points, the deep learning model is relegated to a reserve role in the dual-model approach. The incorporation of static analysis automation into the system streamlines processes and saves valuable time. The static analysis report, which includes the assembly code of the binary file, provides a comprehensive basis for further investigation. In summary, the malware analysis system, employing a dual-model approach, effectively integrates deep learning models and reduces false positives.

The network traffic classifier demonstrates significant potential in improving network security by accurately analysing real-time traffic and detecting a wide range of cyber threats. Using Random Forest as the main model provided strong results, especially with the CIC-IDS 2017 dataset, where it achieved 99.82% accuracy. This is important as CIC-IDS 2017 includes modern attack types like DDoS and brute force, making the model highly relevant for current security challenges. The NSL-KDD and UNSW-NB15 datasets, with their focus on both traditional and next-generation threats, further validated the classifier's versatility.

A key advantage of this system is its modular, containerized design, which allows each component, from traffic capture to classification, to operate independently. This flexibility ensures scalability and efficient deployment in diverse network environments. The use of containers streamlined the workflow, making it easier to handle large volumes of traffic while maintaining high detection accuracy. Simulated attack testing using various penetration tools confirmed the model's effectiveness, successfully identifying attacks such as DDoS and infiltration. However, like any machine learning-based system, its performance will depend on continuous updates and retraining as new threats emerge. Overall, this research highlights the potential of AI-driven network traffic classification in providing real-time defence against evolving cyber threats.

The web intrusion detection system (WIDS) effectively uses Isolation Forest for real-time anomaly detection in HTTP logs. Unique features were engineered based on HTTP log files, further augmenting the system's capabilities and contributing to its uniqueness. By optimizing hyperparameters like contamination levels, the system strikes a balance between detecting attacks and reducing false positives. The chosen contamination level of 0.01 showed the best results, offering sensitivity to threats without overwhelming false alarms. Real-time testing in a cloud environment proved the system's effectiveness in identifying attacks, such as a simulated directory traversal. The lightweight agent and cloud setup ensured efficient data collection and processing without straining servers. Although the system performs well, defining 'normal' behaviour remains a challenge. Continuous monitoring and updates are essential to keep the model relevant in detecting new threats. This research demonstrates how AI-powered systems can enhance web security in dynamic environments.

To overcome bottlenecks and scale the system across large and complex cloud environments, the deployment techniques outlined offer a robust solution. For the Network Traffic Classifier, horizontal scaling with a load balancer is key. As the traffic increases, additional VM instances can be dynamically added to the pool, distributing the workload across multiple instances. For example, in a scenario where a large enterprise's network sees a sudden spike in traffic, the load balancer can ensure that no single VM is overwhelmed, preventing latency or data loss. This not only ensures efficient processing but also maintains real-time attack detection capabilities across distributed systems. In the case of Web Intrusion Detection, vertical scaling is more suitable due to the increasing complexity of web server logs. By allocating a dedicated VM to each web instance, resource allocation can be fine-tuned to meet the specific demands of each instance. For example, high-traffic websites can benefit from VMs with larger CPU and memory allocations, ensuring the anomaly detection system continues functioning efficiently without delays or resource contention. For the Malware Analysis application, employing a load balancer to distribute file uploads across multiple containers can address the challenge of simultaneously handling large volumes of files. Additionally, dedicating a separate node for API handling isolates the file classification process from the user-facing API, improving system responsiveness and scalability. This separation of concerns allows the system to handle thousands of concurrent file uploads or API requests without performance degradation, ensuring that even in complex environments with heavy workloads, the system remains efficient and responsive. This modular architecture of containerised applications effectively addresses scalability and administration challenges, showcasing the versatility of container technology in deploying AI/ML applications. Deploying various AI/ML applications in containers with different architectures underscores the flexibility and suitability of this deployment method.

While initial results are promising, continuous model updates and rigorous testing are imperative to ensure effectiveness against evolving cyber threats. Moreover, the scalability of the designed system allows for the integration of new AI/ML models, expanding its capabilities in various cybersecurity areas. Deploying all models within a designated subnet enhances security and streamlines administration, with different cloud providers offering diverse solutions for AI/ML deployment. A summary of the proposed AI-enabled system contribution to Cyber Incident Response is in Table X.

TABLE X: Summary of our AI-Enabled System contribution to Cyber Incident Response

| Security Problem | Application | Algorithm | Dataset | Contribution |
|---|---|---|---|---|
| Attack Detection | Network Traffic Classifier | Random Forest | **NSL-KDD, UNSW-NB15, CIC-IDS-2017** | **End-to-end Solution:** An end-to-end solution for network traffic classification encompassing feature extraction from PCAP and the classification model itself. **Automation:** Time and labour-efficient Automation of feature extraction from PCAP files by implementing a feature engineering container. **Real-time Analysis:** Real-time analysis allows an immediate detection and response to network threats. |
| Anomaly Detection | Web Intrusion Detection | Isolation Forest | Private Dataset | **Distributed Data Collection:** The deployment of lightweight agents on individual web servers for efficient log collection and forwarding to the central Web Intrusion Detection System (WIDS). This distributed data collection method minimises the impact on individual web servers while facilitating real-time analysis. Dynamic **Contamination Ratio Determination:** Methodology to optimise the model's sensitivity to anomalies by dynamically determining the contamination ratio for anomaly detection models by simulating attacks and observing the model's detection rate. **Threshold-Based Alarm Triggering:** Threshold-based approach to trigger alarms only when the number of detected anomalies exceeds the preset value, and this minimises unnecessary alerts (false positives) by prioritizing significant deviations from normal behaviour. **Feature Engineering from HTTP Logs:** Novel approach to feature engineering tailored to HTTP server logs, resulting in the creation of five features that capture relevant aspects of web server activity and enable the detection of breaches with high accuracy and efficiency. |
| Malware Detection | Malware Analysis System | Hybrid (RF & Keras) | VirusTotal | **Combined Model Architecture:** Hybride architecture consisting of a random forest model and a secondary Keras TensorFlow models mitigating false positives by leveraging the strengths of both models. **Utilization of LSTM Network for Deep Learning:** Long Short-Term Memory (LSTM) network, a type of Recurrent Neural Network (RNN) suitable for analysing sequential data like text, allows the model to effectively capture temporal dependencies in the data, enhancing its ability to detect subtle patterns indicative of malicious behaviour. **Real-Time Analysis with User Interface:** user-friendly web interface that allows users to upload executable files for real-time malware analysis. The process is streamed including file features analysis, maliciousness prediction, and detailed analysis report generation. |

## VII. LIMITATIONS AND FUTURE WORK

The system discussed in this paper presents certain limitations that highlight areas for future improvement. One significant limitation is that while the system has been tested against a range of known attack types, it has yet to be rigorously evaluated under more comprehensive and diverse real-world threat scenarios. This includes handling the uncertainties associated with dynamic and evolving cyber threats, which may behave unpredictably compared to controlled environments. Additionally, when it comes to anomaly detection in web applications, distinguishing between what constitutes normal behaviour and what qualifies as an anomaly becomes increasingly complex as the volume of data grows. The detection of outliers, especially when processing large datasets, remains a challenge. Another area of concern is the occurrence of false positives, where the system incorrectly flags benign activities as malicious, which can lead to unnecessary alarms and impact the system's overall accuracy.

Looking forward, there are several avenues for future work to address these limitations. First, a more comprehensive set of attack scenarios needs to be incorporated into testing to enhance the system's robustness against real-world cyber threats. Continuous training of models using honeypot data will be crucial for keeping the system adaptable to new and emerging attack patterns. To support this, the exploration of TPU-powered VM instances will enable more efficient training of deep learning models within the DFIR environment, overcoming current computational constraints. Moreover, automating the deployment process will enable seamless system deployment across any cloud environment—whether AWS, Azure, or GCP enhancing the system's flexibility and responsiveness. The use of this system design with multiple virtual private networks (VPNs) and Docker containers provides full control over the design while also supporting a multi-cloud approach, which is a direction for future research.

In addition to the outlined future work, another promising direction is the exploration of how models can be tailored to meet specific organisational needs. For instance, organizations could benefit from models specifically trained for phishing detection, which could analyse incoming emails in real-time to identify potential phishing attempts based on known patterns and emerging tactics. Similarly, models could be developed to detect insider threats by analysing user behaviour across systems. This flexible, model-driven approach would allow organizations to select and customise AI/ML solutions to address their unique cybersecurity challenges.

Another distinctive feature to be investigated is the capability to update models using honeypot data collected from the T-Pot platform deployed within the honeypot environment, as illustrated previously in Figure 3. Within this environment, a storage bucket is established to facilitate the transmission of essential log data from the honeypot. A virtual machine instance, situated in the research and development subnet of the DFIR environment, gathers data and performs data processing and feature engineering. The processed data is subsequently labelled and uploaded to the training data bucket. Leveraging this training data, machine learning models are trained within the cloud environment using TPU-powered VMs and tested, which results in a new, up-to-date version of the model that is then pushed to the system. The type of training data extracted is tailored to the functionality of each model. This paper presents the concept of the honeypot subsystem, which warrants further investigation in future studies. By continuously exposing the system to potential attack vectors through multiple honeypot daemons and consistently gathering data to train the machine learning models, this approach ensures that the system remains at the forefront of threat detection and incident response. Moreover, it enhances the system's adaptability to the ever-evolving cybersecurity landscape.

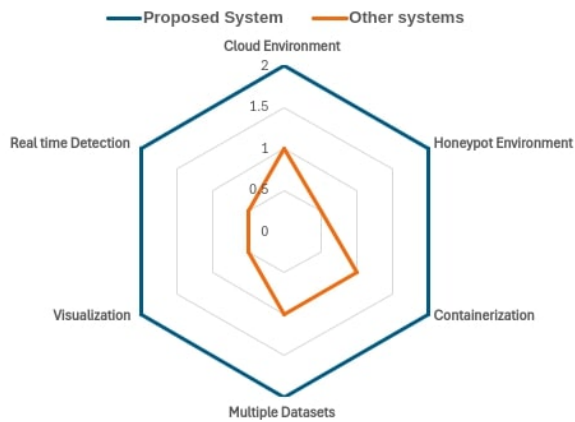

Fig. 9: Proposed System Performances [45] Comparison with Other Related Works.

Finally, expanding the use of deep learning across all applications within the system will further strengthen its ability to handle complex and diverse cybersecurity challenges [46]. Since neural networks have shown tremendous results, as indicated in malware detection applications, it is highly suitable to introduce neural networks for the network traffic classifier and web intrusion anomaly detector and also to adapt a dual model approach to reduce false positives. For a visualised perspective highlighting our system contribution, we elaborated Figure 9 which presents a comparative spider radar map in terms of novelty, completeness, and innovativeness. The proposed system offers a comprehensive solution that surpasses existing approaches and systems in several key dimensions.

## VIII. CONCLUSION

In this paper, we presented a comprehensive exploration of the practical use of Artificial Intelligence (AI) techniques in the context of cyber security, focusing on their integration into incident response systems within cloud environments. Through the development and deployment of a cyber threat defence system, which included a network traffic classifier application, malware analysis application, and web intrusion detection system, we demonstrated the practical application of AI/ML in enhancing cyber security capabilities. Our research highlighted the potential of AI/ML to address emerging cyber threats and improve the efficiency and effectiveness of cyber incident investigations. By deploying the system on both Google Cloud and Microsoft Azure platforms, we showcased the scalability and versatility of AI-powered cyber security solutions in cloud environments. Through testing on multiple datasets like NSL-KDD, CIC-IDS 2017, UNSW-NB15, and VirusTotal samples, Random Forest models consistently achieved high accuracy in detecting a range of cyber threats, while deep learning models offered additional precision despite their higher computational cost. The modular, containerized architecture of the system ensures efficient deployment and scalability across cloud environments, allowing for real-time traffic analysis and threat detection.

Furthermore, the integration of a T-pot for continuous development of models and the ELK Stack for log gathering and visualisation, emphasising the importance of comprehensive data analysis and visualisation in cyber security operations. These tools, combined with AI/ML techniques, offer a holistic approach to cyber security that enables proactive threat detection and rapid incident response. Overall, our findings underscore the critical role of AI/ML in modern cyber security and highlight the need for continued research and development in this field. As cyber threats continue to evolve, leveraging advanced technologies such as AI and ML will be essential for staying ahead of adversaries and ensuring the security of digital assets and infrastructure.

### FOOTNOTES

*Ethical Approval*

This research was deemed not requiring ethical approval.

*Funding*

The APC and Open Access fees for this work are funded by the University of Liverpool.

*Availability of data and materials*

Code generated for this research is detailed in [45] and made publicly available on GitHub **https://github.com/Ashfaaq98/Periscope-AI**.

*Competing Interests*

The authors declare that they have no known competing interests or personal relationships that could have appeared to influence the work reported in this paper.

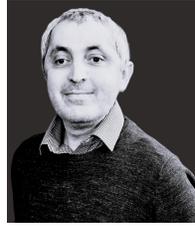

Dr Ayman El Hajjar is a Senior Lecturer and head of the Cyber Security Research Group at the School of Computer Science and Engineering at the University of Westminster. He also leads the Cyber Security and Forensics BSc program. While leading the university's Cyber Security research group, the group has successfully secured substantial funding. This funding supports their extensive research in IoT security, data integrity, Incident Response, and cryptographic solutions. His primary research focus centres on the security of the Internet of Things (IoT) and Smart Cities. Within these domains, Dr El Hajjar and his PhD students are exploring the use of various cryptographic models to secure communication of IoT devices both in transit and at rest, specifically focusing on data confidentiality and integrity and enabling secure key distribution among devices. Dr El Hajjar is also interested in addressing the human factor in cybersecurity He actively seeks solutions that lessen the burden on end-users, advocating for a paradigm shift towards more accessible and user-friendly security practices. Dr El Hajjar is also a member of the UK Cyber Security Council ethical committee, ensuring the ethical practices of all the council's members are maintained to the highest standard. He has published numerous research papers and edited books in the field of cybersecurity. His publications cover many topics, including IoT security, data integrity, cryptography, and smart city technologies. Dr El Hajjar is also a reviewer for top cyber security and cryptography journals.

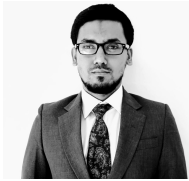

Mr Mohammed Ashfaaq Farzaan is an R&D Lead Engineer working on Applied AI at PRIAM CYBER AI Limited UK. Mr Ashfaaq Farzaanholds a Bachelor's Degree in Information Technology with a specialisation in Computer Systems and Network Engineering, and an MSc in Cyber Security with Distinction from University of Westminster UK. He is a Multi-Certified Cyber Security Experts holding many Certification such as ISC2 Certified in Cyber Security (CC), Blue Team Juniot Analyst (BTJA) and has collaborated with various professionals from the industry on different projects, including Autonomous Incident Response, LLM-based Threat Hunting, and Malware Analysis. His interests span multiple disciplines, including Cybersecurity, Digital Forensics, Generative AI, Large Language Models, Security Operations Center (SOC), and IoT/Cloud.

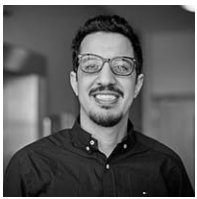

Dr Mohamed Chahine Ghanem is an Associate Professor in Cyber Security and AI, the Director of the Cyber Security Research Centre/ London Metropolitan University, and Associate Professor in Cyber Security within the Department of Computer Science / the University of Liverpool. Dr Ghanem holds an Engineering Degree in Computer Sciences, MSc in Digital Forensics with Distinction and a PhD in Cyber Security from the City, University of London. He is a Senior Fellow of HEA and holds a PGDip in Security Studies. Dr Ghanem is an IEEE member and a professional member of the British Computer Society achieved many certificates, such as GCFE, CISSP, ACE, XRY and CPCI with over 15 years experience at senior management level in the field of cybersecurity, digital forensics and incident investigation in law-enforcement and world-leader corporates. His research focuses on applying AI to solve real-world cybersecurity and digital forensics problems and published numerous research papers in the world's top cybersecurity journals. He delivered several keynote lectures at different prestigious venues. Dr Ghanem is Academic Editor of ACM Digital Threats: Research and Practice journal and Senior Advisor at KROLL LLC.

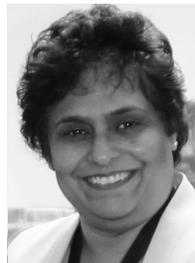

Dr. Deepthi N. Ratnayake is a Principal Lecturer in Cyber Security at the University of Hertfordshire, UK, an active contributor to the Cybersecurity and Computing Systems Research Group and Programme Leader for the Modular Masters Programme in Computer Science. With over 30 years of experience spanning industry, defense, and academia, Dr. Ratnayake brings significant expertise in cybersecurity, networking, and information systems management. Her previous role as Head of a defense network has provided her with both technical depth and strategic insight that she applies to her current work. Dr. Ratnayake holds a PhD specialising in Probe Request Attack Detection in Wireless LANs using Intelligent Techniques, and her research interests encompass Intrusion Detection and Prevention through intelligent techniques, Security in Cloud Computing, Software Defined Networks, and Information Security Management and Compliance. She is an Associate Editor for the Information Security Journal: A Global Perspective, serves as a Cyber Security Columnist, and is an active, longstanding member of the Information Security Specialist Group of the British Computer Society (BCS-ISSG). A passionate advocate for addressing real-world cybersecurity issues, Dr. Ratnayake has presented extensively at academic and professional venues, authored high-impact journal articles, and contributed to both national and international conferences. Her work continues to advance state-of-the-art cybersecurity practices and solutions.